\documentclass[11pt]{article}
\usepackage{amssymb}
\usepackage{subfig}
\usepackage{color}
\usepackage{epsfig,amssymb,amsfonts,amsmath,graphicx,dsfont,cite,xfrac}
\usepackage{authblk}
\definecolor{mygray}{gray}{0.5}
\usepackage{cite}
\usepackage[colorlinks=true,linkcolor=blue,citecolor=red]{hyperref}
\usepackage{multirow}

\usepackage{tikz,pgfplots}
\usetikzlibrary{decorations.pathreplacing}
\usetikzlibrary{shapes.multipart}
\usetikzlibrary{positioning}
\usetikzlibrary{matrix}
\usetikzlibrary{shapes,calc,arrows}

\newcommand{\be}{\begin{equation}}
\newcommand{\ee}{\end{equation}}
\newcommand{\bea}{\begin{eqnarray}}
\newcommand{\eea}{\end{eqnarray}}

\title{SUSY transformation as the coupler of non-interacting systems}

\author{V. Jakubsk\'y$^{1}$}

\affil{$^1$Nuclear Physics Institute, Czech Academy of Science, 250 68 \v{R}e\v{z}, Czech Republic\\\vspace{3mm}jakubsky@ujf.cas.cz}

\begin{document}
\maketitle
\begin{abstract}
Quasi-one-dimensional chains of atoms can be effectively described by one-dimensional Dirac-type equation. Crystal structure of the chain is reflected by pseudo-spin of the quasi-particles. In the article, we present a simple framework where supersymmetric transformation is utilized to generate an interaction between two, initially non-interacting systems described by pseudo-spin-one Dirac-type equation. In the presented example, the transformation converts two asymptotically non-interacting atomic chains into a saw chain locally. The model possesses a flat band whose energy can be fine-tuned deliberately. 

\end{abstract}

\section{Introduction}
The supersymmetric (also called Darboux) transformation has been a significant tool in the study of differential equations for a long time \cite{Darboux}. Its primary function is to transform one differential equation into another while preserving the solvability of the original equation. In the realm of quantum physics, the transformation has been revitalized in the context of supersymmetric quantum mechanics, where it plays a crucial role in forming supercharges and intertwining operators that map between superpartner Hamiltonians~\cite{Witten},  \cite{COOPER}. 

This technique has also found applications in condensed matter physics, particularly in systems described by low-dimensional Dirac equations.
Two main approaches have been adopted in this area. The first approach takes advantage of the fact that squaring a $2\times2$ one-dimensional Dirac operator with a minimally coupled magnetic field results in a diagonal Schr\"odinger operator. By applying the supersymmetric (susy) transformation to Schr\"odinger operators, the new Schr\"odinger operator's square root can be identified with the Dirac Hamiltonian of the transformed system \cite{KuruNegroNieto,MidyaFernandez,Jakubsky13}, \cite{Jahani,Phan20,CastilloCeleita20,Fernandez20,Fernandez20b,SchulzeRoy,Celeita,Concha-Sanchez,ContrerasNegro}.

In the second approach, the susy transformation is applied directly to Dirac operators. This method allows for the identification of solvable configurations involving effective interactions that are different from magnetic fields. This approach has been explored for stationary one-dimensional $2\times2$ Dirac equations \cite{Samsonov} and non-stationary equations \cite{Samsonov2}. There have been various extensions of the susy transformation proposed for Dirac operators in polar coordinates \cite{Sch22}, higher dimensions \cite{Schulze1,Iof22}, and higher spin systems~\cite{Schulze13}. Further generalizations of the intertwining relations for Dirac operators have also been investigated~\cite{Ioffe}. 

Current experimental advances make it possible to compose quasi-one-dimensional atomic chains described by pseudo-spin-1 Dirac equation that possess flat band in the energy spectrum.  Let us mention e.g. stub, diamond (rhombic), Creutz, or fishbone lattices \cite{Real,Huda2,Travkin}.
The low-energy spectrum of pseudo-spin-1/2 and pseudo-spin-1 Dirac materials can be substantially different. In the first case, the two energy bands form the Dirac cone with a possible gap between positive and negative energies. In case of pseudo-spin-1 systems, the Dirac cone is there as well, but it comes accompanied by a third energy band. In some crystals, the third band is independent on momentum, it is perfectly flat. The flat-band is associated with vanishing group velocity and macroscopical degeneration of eigenstates. It was observed experimentally in optical lattices \cite{Travkin,Mukherjee17,Baboux}. 
Transport or topological spectral properties of the quasi-one-dimensional lattices, for instance of decorated or bearded SSH chains, were studied recently in \cite{Betancur1}, \cite{SSHJakubsky}, \cite{Betancur2}, \cite{Verma}.

We will show how susy transformation can be utilized in designing solvable model of quasi-one-dimensional chain described by pseudo-spin-1 Dirac Hamiltonian. By construction, the model will possess flat band in the spectrum. We will discuss physical relevance of the resulting Hamiltonian operators in the context of quasi-one-dimensional crystals formed by two parallel chains of atoms. We will show that susy transformation can serve as the generator of coupling between two parallel chains of atoms, forming a "saw chain" locally.

The work is organized as follows: the quasi-one-dimensional chain of atoms forming the saw chain will be discussed in detail in the next section. In the third section, susy transformation is employed to generate coupling between two initially non-interacting systems. General formula is provided for the Hamiltonian that can be obtained this way. It is compared with the low-energy operator of saw chain. In the fourth section, two explicit examples are discussed where coupled systems with flat-band in their spectrum are constructed. The work uses and extends the results of \cite{SSHJakubsky}  where susy transformation was employed in spectral engineering of stub lattice. The last section is left for discussion. 

\section{The saw chain}
Let us consider a quasi-one-dimensional crystal that has three atoms in the elementary cell. The atoms are arranged into a triangle that is periodically replicated along the $x$-axis. The coupling among the triangles is represented by the hopping parameter $\tilde{t}_{AB}$, whereas the interaction among the atoms $A$, $B$, and $C$ within the elementary cell is reflected by the hopping parameters $t_{AB}$, $t_{AC}$, and $t_{BC}$, see Fig.1. The tight-binding Hamiltonian for the electrons on the lattice reads as follows
\begin{align}
H&=\sum_{n=-\infty}^{\infty}\left(t_{AB}A^{\dagger}_nB_n+\tilde{t}_{AB}A^{\dagger}_nB_{n-1}+t_{AC}A^\dagger_nC_n+t_{BC}B^\dagger_nC_n+h.c.\right)\nonumber\\&+\sum_{n=-\infty}^{\infty}\left(t_{AA}A_n^\dagger A_n+t_{BB}B^\dagger_nB_n+t_{CC}C^\dagger_nC_n\right),
\end{align} 
where $X_j^\dagger$ and $X_j$ are creation and annihilation operators of electrons on the corresponding sites, $X=A,B,C$.
\begin{figure}
	\centering
	\includegraphics[width=0.8\textwidth]{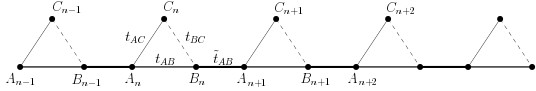}
	\caption{The saw chain. The upper line formed by the atoms $C_j$ corresponds to forms the $C$-chain. The lower line composed of the atoms $A_j$ and $B_j$ forms the $AB$-chain, $j\in\mathbb{Z}$ }
	\label{fig10}
\end{figure}
In the basis of Bloch states with quasi-momentum $k$, the Hamiltonian can be written in the following form
\begin{equation}\label{TBH}
H(k)=\begin{pmatrix}t_{AA}&t_{AB}+\tilde{t}_{AB}e^{-ika}&t_{AC}\\\dots&t_{BB}&t_{BC}\\h.c.&\dots&t_{CC}\end{pmatrix},
\end{equation}
where $a$ is the length of the primitive translation vector.
The saw chain Hamiltonian is a generalization of stub-lattice operator that was discussed e.g. in \cite{SSHJakubsky}.
The spectrum of $H(k)$ has three bands that correspond to the solutions of the secular equation
\begin{equation}\label{det}
\det(H-E(k))=0.
\end{equation}
All of the solutions depend on the momentum $k$ in general, $E_{j}=E_{j}(k)$, $j=1,2,3$. Therefore, there is no flat band for generic values of the hopping parameters $t_{AB}$, $\tilde{t}_{AB}$, $t_{AC}$, $t_{BC}$, $t_{AA}$, $t_{BB}$, and $t_{CC}$. Nevertheless, the flat band can be recovered provided that one of the on-site energies, for instance $t_{CC}$, is fine-tuned conveniently. Indeed, one can match the determinant with a polynomial of the third order,
\begin{equation}
\det(H-E(k))=(E-a_2)(E^2+a_1E+a_0).
\end{equation}
The parameter $a_0$ can be fixed such that it depends on the hopping parameters $t_{AB}$, $\tilde{t}_{AB}$, $t_{AC}$, $t_{BC}$, $t_{AA}$, $t_{BB}$, and $k$, while $a_1$, $a_2$ as well as $t_{CC}$ can be fixed such that they depend solely on the hopping parameters $t_{AB}$, $\tilde{t}_{AB}$, $t_{AC}$, $t_{BC}$, $t_{AA}$, $t_{BB}$ and are independent of $k$. It is rather straightforward to find the explicit forms of $a_0$, $a_1$, $a_2$ and $t_{CC}$. However, the expressions are rather lengthy so that we refrain from presenting them explicitly here.  The parameter $a_2$ fixes the energy of the flat band, see Fig.\ref{fig10a} for illustration. 

\begin{figure}
	\centering
	\includegraphics[width=0.4\textwidth]{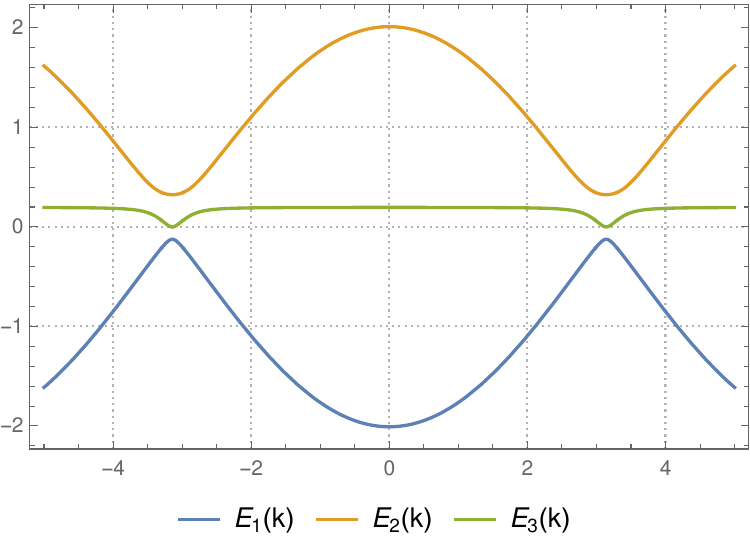} 
	\includegraphics[width=0.4\textwidth]{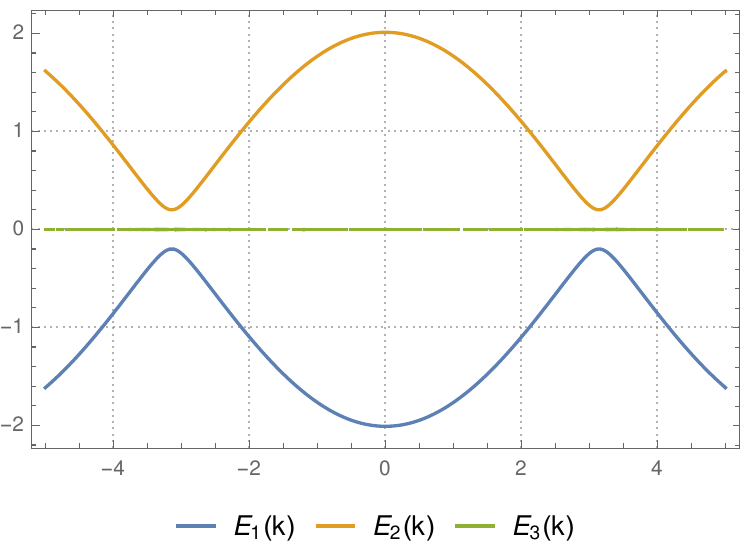} 
	\caption{The energy bands $E_j(k)$, $k=1,2,3$, as the solutions of 
		(\ref{det}). We set $a=1$, $t_{AA}=0$, $t_{BB}=0$, $t_{AB}=\tilde{t}_{AB}=1$, $t_{AC}=0.2$, $t_{BC}=0.01$ and left: $t_{CC}=0.2$, right: $t_{CC}=1/500$, $a_2=0$ in this case. There is quasi-momentum $k$ on the horizontal axis whereas the vertical axis corresponds to the energy.}
	\label{fig10a}
\end{figure}

When the hopping parameters $t_{AB}$ and $\tilde{t}_{AB}$ are approaching the same value, two energy bands come close together in the corners of the first Brillouin zone, forming there the Dirac cones. The two Dirac cones are equivalent as they can be connected by the primitive translation vector in the reciprocal lattice. Expanding the Hamiltonian (\ref{TBH}) in the vicinity of these point provides us with the following Dirac-type operator
\begin{equation}
\tilde{H}_D=\begin{pmatrix}t_{AA}&t_{AB}-\tilde{t}_{AB}+ia\,\tilde{t}_{AB}k_x&t_{AC}\\t_{AB}-\tilde{t}_{AB}-ia\,\tilde{t}_{AB}k_{x}&t_{BB}&t_{BC}\\t_{AC}&t_{BC}&t_{CC}\end{pmatrix},
\end{equation} 
where $k_x=-i\partial_x$. 
We make an additional transformation $U=\mbox{diag}(1,i,i)$ to turn the kinetic term into the more convenient form,
\begin{equation}\label{DiracSawChain}
H_D=U\tilde{H}_DU^{-1}=\begin{pmatrix}t_{AA}&-i\,(t_{AB}-\tilde{t}_{AB})-i\,\tilde{t}_{AB}a\,\partial_x&-i\,t_{AC}\\i\,(t_{AB}-\tilde{t}_{AB})-i\,\tilde{t}_{AB}a\,\partial_x&t_{BB}&t_{BC}\\i\,t_{AC}&t_{BC}&t_{CC}\end{pmatrix}.
\end{equation}
We will fix $a=1$ if not specified otherwise.

In the following section, we will show how Darboux transformation can be used to generate a solvable model of atomic chains where the hopping parameters get position-dependent.

\section{Coupling via Darboux transformation }
Let us consider the following one-dimensional pseudo-spin-1 Dirac Hamiltonian
\begin{equation}\label{H}
H=\begin{pmatrix}m(x)+v(x)&-i\partial_x-iA(x)&0\\-i\partial_x+i A(x)&-m(x)+v(x)&0\\0&0&\lambda\end{pmatrix},
\end{equation}
where $m(x)$, $v(x)$, $A(x)$, and $\lambda$ are real-valued so that $H$ is hermitian. The Hamiltonian represents a quantum system where two subsystems coexist without mutual interaction. In one of them, dynamics is driven by the $2\times2$ diagonal block operator while the dynamics in the second one is frozen as the energy operator is constant.
Such system can correspond to a couple of parallel chains of atoms. The two systems can correspond to two parallel atomic chains. The atoms $A$ and $B$ in the $AB$-chain are mutually interacting, being described by the $2\times2$ diagonal block in (\ref{H}). The atoms $C$ in the $C$-chain do not interact neither with their neighbor atoms nor with the atoms from the first chain, see Fig. 3 for illustration.

Let us suppose that the eigenvectors of $H$ are known. Let us fix two eigenvectors  corresponding to the eigenvalue $\lambda$ and one eigenvector corresponding to a constant $\epsilon\in\mathbb{R}$. We use them to compose a matrix $U$ such that it satisfies the following equation,  
\begin{equation}\label{U}
U=\begin{pmatrix}i\psi_0&i\psi_1&i\psi_2\\\phi_0&
\phi_1&\phi_2\\0&\xi_1&\xi_2\end{pmatrix},\quad HU=U\begin{pmatrix}\epsilon&0&0\\0&\lambda&0\\0&0&\lambda\end{pmatrix},\quad \epsilon,\lambda\in\mathbb{R},
\end{equation}
where the functions $\psi_a$, $\phi_a$, $a=0,1,2$ and $\xi_1$ and $\xi_2$ are fixed to be real-valued. 
The columns of $U$ are formed by the eigenvectors corresponding to the eigenvalues $\epsilon$ or $\lambda$, respectively. It is worth mentioning that the functions $\xi_1$ and $\xi_2$ are arbitrary. It follows from the fact that $(0,0,\xi(x))^t$ for any $\xi(x)$ is the eigenvector of $H$ corresponding to the eigenvalue $\lambda$.
The matrix $U$ can be used to define the new Dirac-type operator $\tilde{H}$ and an operator $L$,
\begin{align}\label{H1gen}
\tilde{H}&=-i\gamma\partial_x+\tilde{V},\quad\tilde{V}=V-i[\gamma, (\partial_xU)U^{-1}],\quad \gamma=\begin{pmatrix}0&1&0\\1&0&0\\0&0&0\end{pmatrix},\\
L&=U\partial_xU^{-1}.
\end{align}
The Hamiltonians $\tilde{H}$, $H$ and the operator $L$ satisfy the following operator relation,
\begin{equation}\label{Lgen}
LH=\tilde{H}L,
\end{equation}
see \cite{Samsonov} for more details.
The operator $L$ represents the Darboux transformation. It transforms the eigenstates of $H$ into the eigenstates of $\tilde{H}$ with exception of the states that form the kernel of $L$. Indeed, there holds
\begin{equation}
(H-\epsilon)\Psi=0\Longrightarrow (\tilde{H}-\epsilon)\tilde{\Psi}=0,\quad \tilde{\Psi}=L\Psi.
\end{equation} 
By construction, the matrix $(U^{-1})^\dagger$ satisfies 
\begin{equation}
\tilde{H}(U^{-1})^\dagger=(U^{-1})^\dagger\begin{pmatrix}\epsilon&0&0\\0&\lambda&0\\0&0&\lambda\end{pmatrix},
\end{equation}
see \cite{Samsonov}. Therefore, the columns of the matrix $(U^{-1})^\dagger$ can be identified with the eigenvectors of the new Hamiltonian $\tilde{H}$. They can be square-integrable and form the bound states of the new system. 

The new potential $\tilde{V}$ is not hermitian in general. Nevertheless, the function $\xi_1$ can be fixed in order to recover hermiticity of $\tilde{ V}$. It is sufficient to fix $\xi_1$ in the following manner, see \cite{SSHJakubsky},
\begin{equation}\label{xi1}
\xi_1=\xi_2\left(c_1-\int\frac{(\epsilon-\lambda)(\phi_2\psi_1-\phi_1\psi_2)}{\xi_2^2}dx\right)=\xi_2\left(c_1-\int\frac{(\epsilon-\lambda)\,W_0}{\xi_2^2}dx\right).
\end{equation}
Here, $c_1$ is a real integration constant. The quantity $W_0\equiv\phi_2\psi_1-\phi_1\psi_2$ is a real constant as well as there holds $\partial_xW_0=0$. It follows from the fact that $(\psi_1,\phi_1,0)^t$ and $(\psi_2,\phi_2,0)^t$ are eigenvectors of $H$ corresponding to the same eigenvalue.

As it was shown in \cite{SSHJakubsky}, the new potential $\tilde{V}$ defined in (\ref{H1gen}) has the following form 
\begin{equation}\label{Htight}
\tilde{H}=\begin{pmatrix}0&-i\partial_x&0\\-i\partial_x&0&0\\0&0&0\end{pmatrix}+\begin{pmatrix}\tilde{v}_{11}+v&-i\, \tilde{v}_{12}&-i \,\tilde{v}_{13}\\i \tilde{v}_{12}&-\tilde{v}_{11}+v&\tilde{v}_{23}\\
i\, \tilde{v}_{13}&\tilde{v}_{23}&\lambda\end{pmatrix}, 
\end{equation}
where
\begin{align}
\tilde{v}_{12}&=-A+\left(\epsilon-\lambda\right)\frac{\psi_0\,(\xi_2\psi_1-\xi_1\psi_2)-\phi_0\,(\xi_2\phi_1-\xi_1\phi_2)}{\psi_0\,(\xi_2\phi_1-\xi_1\phi_2)-\phi_0\,(\xi_2\psi_1-\xi_1\psi_2)},\\
\tilde{v}_{13}&=\frac{(\epsilon-\lambda)\,\psi_0\,(\phi_1\psi_2-\phi_2\psi_1)}{\psi_0\,(\xi_2\phi_1-\xi_1\phi_2)-\phi_0\,(\xi_2\psi_1-\xi_1\psi_2)}%=-\frac{(\epsilon-\lambda)\,\psi_0\,W_0}{\psi_0\,(\xi_2\phi_1-\xi_1\phi_2)-\phi_0\,(\xi_2\psi_1-\xi_1\psi_2)}
,\\
\tilde{v}_{23}&=-\frac{(\epsilon-\lambda)\,\phi_0\,(\phi_1\psi_2-\phi_2\psi_1)}{\psi_0\,(\xi_2\phi_1-\xi_1\phi_2)-\phi_0\,(\xi_2\psi_1-\xi_1\psi_2)},\\
\tilde{v}_{11}&=-m+(\epsilon-\lambda)\frac{\psi_0\,(\xi_2\phi_1-\xi_1\phi_2)+\phi_0\,(\xi_2\psi_1-\xi_1\psi_2)}{\psi_0(\xi_2\phi_1-\xi_1\phi_2)-\phi_0(\xi_2\psi_1-\xi_1\psi_2)}.\label{vij}
\end{align}

The operator $\tilde{H}$ represents a fundamentally different physical reality. Now, the subsystems are interacting via $\tilde{v}_{13}$  and $\tilde{v}_{23}$. When compared with the Dirac operator of the saw chain (\ref{DiracSawChain}), we notice that the susy transformation generates coupling between the $C$-chain and the $AB$-chain, forming the saw lattice locally at least, see Fig.\ref{fig10} for illustration. In the following section, we will illustrate the framework on an explicit example.

\begin{figure}
	\centering
	\includegraphics[width=0.8\textwidth]{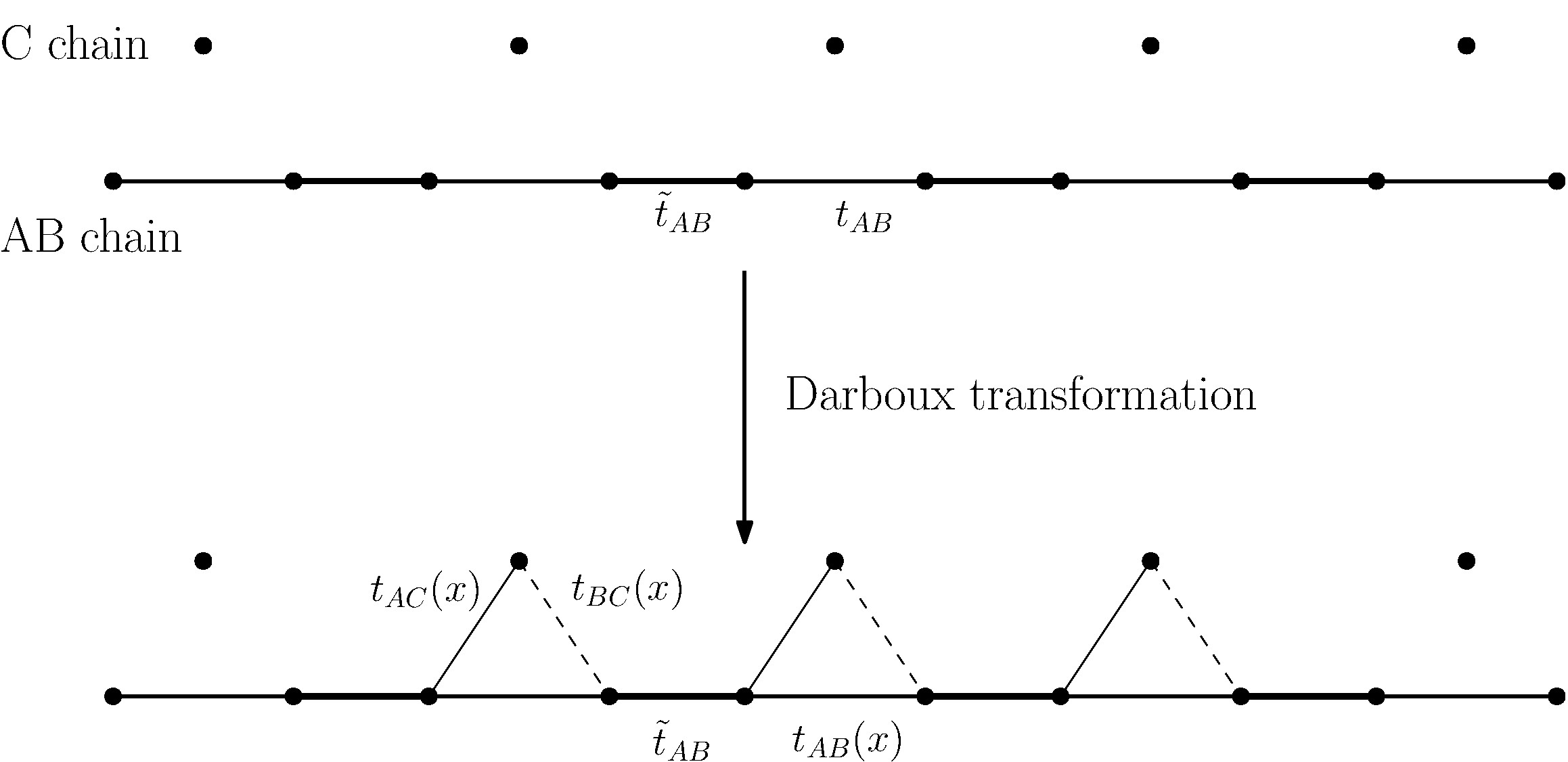}
	\caption{The Hamiltonian (\ref{H}) describes two parallel non-interacting chains of atoms, $C$ chain and $AB$ chain. Susy transformation generates the new potential terms (\ref{vij}) that correspond to spatially dependent couplings $t_{AC}(x)$ and $t_{BC}(x)$ between the two atomic chains. It can also make the coupling $t_{AB}$ inhomogeneous.}
	\label{fig10}
\end{figure}

\section{Flat band of tunable energy in the gap}
Let us take the initial Hamiltonian $H$ in (\ref{H}) in the following simple form 
\begin{equation}
H=
\begin{pmatrix}m&-i\partial_x-iA&0\\-i\partial_x+iA&-m&0\\
0&0&\lambda\end{pmatrix}, \quad m,A,\lambda\in\mathbb{R}.
\end{equation}
Its eigenvectors can be found for any $\epsilon$. We will fix  
\begin{equation}
\epsilon=m.
\end{equation}
The components of the corresponding eigenvector acquire particularly simple form
\begin{equation}\label{psi0}
\psi_0=- m e^{-Ax},\quad \phi_0=\,A \,e^{-A x},\quad (H_{1}-m)(i\psi_0,\phi_0,0)=0.
\end{equation}
We will assume that $|\lambda|\neq |m|$. Then the components $\psi_{1,2}$ can be written in the following manner,
\begin{equation}\label{psi12}
\psi_{1,2}=\frac{\phi_{1,2}'+A\phi_1}{m-\lambda},\quad \phi_1=\phi_2\left(\int\frac{W_0}{\phi_2^2}+c_0\right),
\end{equation}
where $c_0$ is a real constant. It follows from the fact that $\phi_1$ and $\phi_2$ are two solutions of the same differential equation of the second order. We assume that they are linearly independent, i.e. $W_0\neq 0$.

Taking these definitions into account, the explicit form of the matrix $U$ relies on the undefined components $\phi_2$ and $\xi_2$, where $\xi_2$ has to be real but it is arbitrary otherwise.
The potential $\tilde{V}$ is calculated in terms of $U$, see (\ref{H1gen}). Therefore, it is defined in terms of $\phi_2$ and $\xi_2$ as well as the constants $W_0$, $c_0$ and $c_1$, see (\ref{psi0}), (\ref{psi12}) and (\ref{xi1}).  We fix the two functions as follows,
\begin{equation}\xi_2=\phi_2=\cosh(\sqrt{A^2+m^2-\lambda^2}\,x).\end{equation} 
Despite their simple form, the explicit formulas for the components of $\tilde{V}$ are rather large.  
Instead of presenting them explicitly, we prefer to fix the parameters such that the potential reduces into a rather convenient form that preserves enough freedom to fine-tune its key parameters, e.g. asymptotic behavior of the interactions or the energy of the flat-band.

\subsection{Model I}
We fix $A$ and $c_0$ in (\ref{psi0}) and (\ref{psi12}) in the following way
\begin{equation}
A=\sqrt{m(m-\lambda)},\quad c_0=W_0\,\omega+c_1,\quad \omega=-\frac{4m}{\sqrt{m(m-\lambda)}(2m+\lambda)}.
\end{equation}
When we fix $c_0$ as in above, the parameter $W_0$ cancels out in all the formulas. The Hamiltonian $\tilde{H}$ depends on two parameters only, $m$ and $\lambda$.
The components of the potential in $\tilde{H}$, see (\ref{vij}), have the following forms
\begin{align}
\tilde{v}_{12}&=-\lambda\frac{2\sqrt{m(m-\lambda)}\,(1+\mbox{sech}\, 2\kappa x) -\kappa\, \tanh 2\kappa x}{2m-\lambda+4m\,\mbox{sech}\,2\kappa\,x},\nonumber\\ \tilde{v}_{13}&=\frac{\sqrt{m(2m+\lambda)}\,\kappa}{4m+(2m-\lambda)\cosh 2\kappa\, x},\quad \tilde{v}_{23}=\frac{\kappa^2}{4m+(2m-\lambda)\cosh 2\kappa\, x},\nonumber\\
\tilde{v}_{11}&=-\frac{\lambda^2+4m^2\, \mbox{sech}\,2\kappa\, x+2\sqrt{m(m-\lambda)}\kappa\tanh 2\kappa\, x}{2m-\lambda+4m\,\mbox{sech}\,2\kappa\,x},\quad \kappa=\sqrt{(m-\lambda)(2m+\lambda)}.\label{vijcase1}
\end{align}
It is easy to see that the elements of the potential are free of singularities provided that $-2m<\lambda<m$. The functions $\tilde{v}_{13}$ and $\tilde{v}_{23}$ vanish exponentially fast for large $|x|$. The functions $\tilde{v}_{12}$ and $\tilde{v}_{11}$ have the following asymptotic behavior,
\begin{align}
\lim_{x\rightarrow\pm\infty}\tilde{v}_{12}&=\pm\frac{\kappa\,\lambda}{2m-\lambda}-\frac{2\sqrt{m(m-\lambda)}\lambda}{2m-\lambda},\\
\lim_{x\rightarrow\pm\infty}\tilde{v}_{11}&=\mp\frac{2\kappa\,\sqrt{m(m-\lambda)}}{2m-\lambda}-\frac{\lambda^2}{2m-\lambda}.
\end{align}
Comparing the potential terms (\ref{vijcase1}) with the Dirac operator of quasi-particles in the saw chain (\ref{DiracSawChain}), we can see that $\tilde{v}_{13}$ and $\tilde{v}_{23}$ correspond to the inter-chain hoppings $t_{AC}$ and $t_{BC}$, respectively. As these components of the potential disappear exponentially, the coupling between the $AB$-chain and $C$-chain is effectively localized on a finite region only. The interaction $\tilde{v}_{11}$ corresponds to the staggered potential on the $AB$-chain. In this chain, the dimerization pattern gets changed as the sign of $(t_{AB}-\tilde{t}_{AB})\sim \tilde{v}_{12}$ gets altered along the $x-$axis.
The behavior of the system can be fine-tuned by variation of the parameters $m$ and $\lambda$. In Tab.\ref{fig4}a), the interactions $\tilde{v}_{13}$ and $\tilde{v}_{23}$ coincide and $\tilde{v}_{12}=0$, whereas in Tab.\ref{fig4}b), $\tilde{v}_{13}$ and $\tilde{v}_{23}$ differ and $\tilde{v}_{12}$ is inhomogeneous.  

\begin{figure}\begin{center}
		\includegraphics[width=0.5\textwidth]{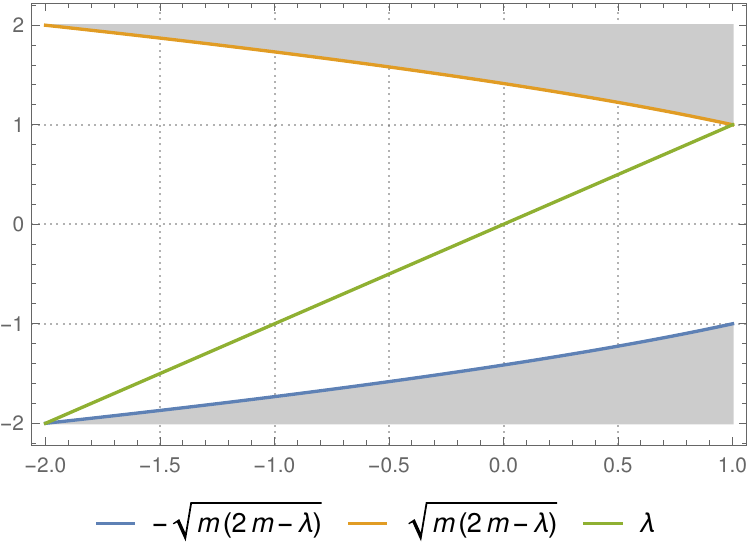}\caption{The two bands of continuum energies with the orange and green threshold enclose the flat band energy $\lambda$. }\label{spectrum2a}\end{center}
\end{figure}

The spectrum of this model has the following form
\begin{equation}
\sigma(\tilde{H})=\left(-\infty,-\sqrt{m(2m-\lambda)}\right]\cup\left[\sqrt{m(2m-\lambda)},\infty\right)\cup\{\lambda\}.
\end{equation}
The thresholds of the continuous bands depend on $\lambda$. We illustrate the dependence on the Fig.\ref{spectrum2a}. It is worth mentioning that $(U^{-1})^{\dagger}$ has two columns that correspond to the bound states of $\tilde{H}$ associated with the energy $\lambda$ of the flat band. Due to the infinite degeneracy of the flat band, the existence of these bound states is rather irrelevant.

\begin{table}
	\begin{center}
		\begin{tabular}{cc}
			\includegraphics[width=0.5\textwidth]{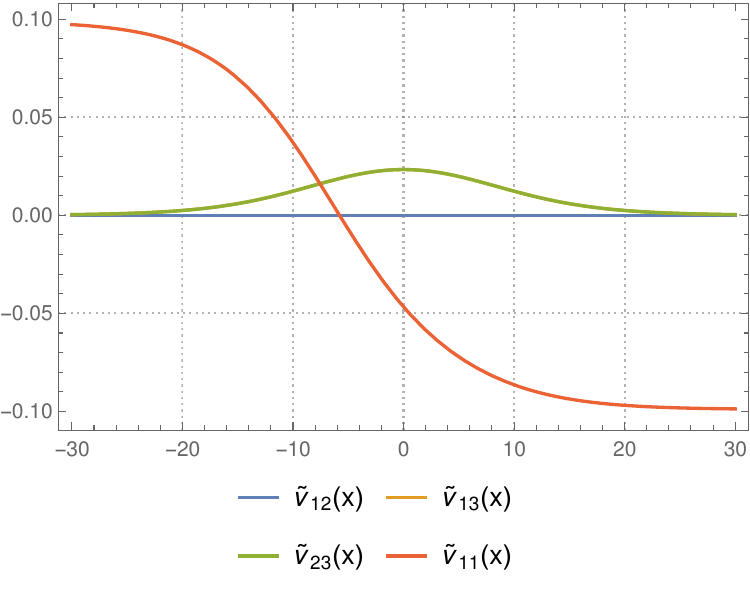}&\includegraphics[width=0.5\textwidth]{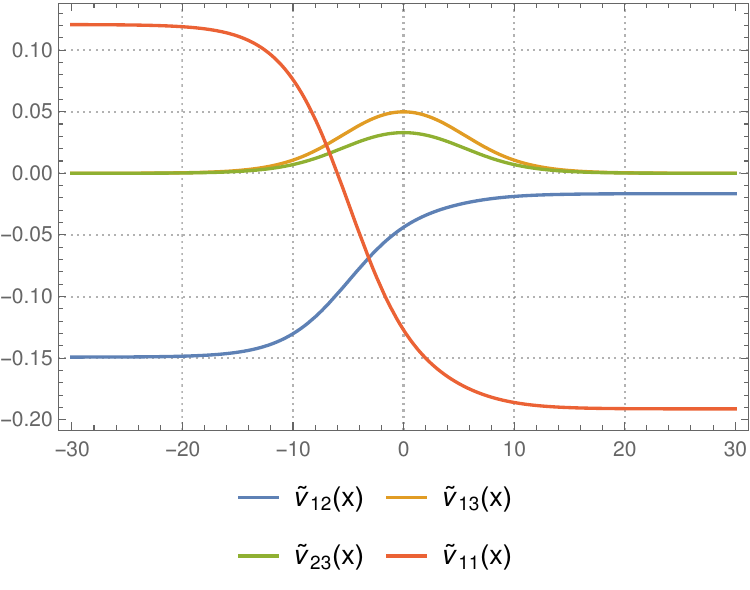}\\
			a)& b)
		\end{tabular}
		\caption{Components of the potential term of $\tilde{H}$, see (\ref{Htight}) and (\ref{vijcase1}). In the figures, we fixed $m=0.07$, $\lambda=0$ on the right and $m=0.16$, $\lambda=0.09$ on the left.   }\label{fig4}
	\end{center}
\end{table}

\subsection{Model II}
Now, we fix the parameters in the following manner,
\begin{equation}
A=m,\quad c_0=W_0\,\omega+c_1,\quad \omega=-\frac{2\,(2m-\lambda)}{(2m^2-\lambda^2)}.
\end{equation}
All other quantities are defined as in the previous model. Likewise in the previous case, $W_0$ cancels out in the final formulas and the Hamiltonian $\tilde{H}$ depends solely on two parameters $\lambda$ and $m$. The components  (\ref{vij}) of $\tilde{V}$ acquire the following form

\begin{align}\label{vmodel2b}
\tilde{v}_{12}&=-\lambda,\quad  \tilde{v}_{13}=\tilde{v}_{23}=\frac{(m-\lambda)\kappa^2}{(2m-\lambda)^2+2(m-\lambda)^2\cosh 2\kappa x},\\
\tilde{v}_{11}&=-\kappa\frac{ (2m-\lambda)\,\kappa\,\mbox{sech}(2\kappa\, x)+2(m-\lambda)^2\tanh\, 2\kappa x}{2(m-\lambda)^2+(2m-\lambda)^2\mbox{sech}2\kappa\, x},\quad \kappa=\sqrt{2m^2-\lambda^2}.
\end{align}
It is straightforward to assess that these components are free of singularities for any values of $m$ and $\lambda$. In order to keep $\kappa$ real, we have to set $\lambda$ such that $\lambda^2<2m^2$.

This model is specific by strictly constant value of $\tilde{v}_{12}$ that has the same magnitude as the energy of the flat-band. As $\tilde{v}_{12}$ corresponds to the relative difference of the hopping parameters $t_{AB}-\tilde{t}_{AB}$, the model allows to fine-tune the dimerization pattern of the $AB$-chain by fixing $\lambda$ appropriately. The components $\tilde{v}_{23}$ and $\tilde{v}_{13}$ coincide and vanish exponentially fast. Therefore, the $C$-chain is coupled to the $AB$-chain only locally, likewise in the previous model.  The mass term $\tilde{v}_{11}$ changes sign asymptotically,
\begin{equation}
\lim_{x\rightarrow\pm\infty}\tilde{v}_{11}=\mp \kappa.
\end{equation}
In Tab.\ref{tab2}, we present the plots of the functions (\ref{vmodel2b}) for different values of $m$ and $\lambda$.

\begin{table}
	\begin{center}
		\begin{tabular}{cc}
			\includegraphics[width=0.4\textwidth]{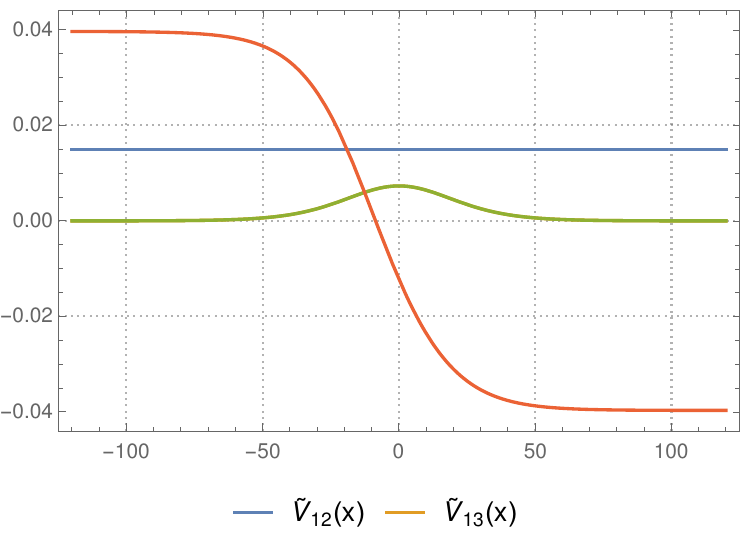}&\includegraphics[width=0.4\textwidth]{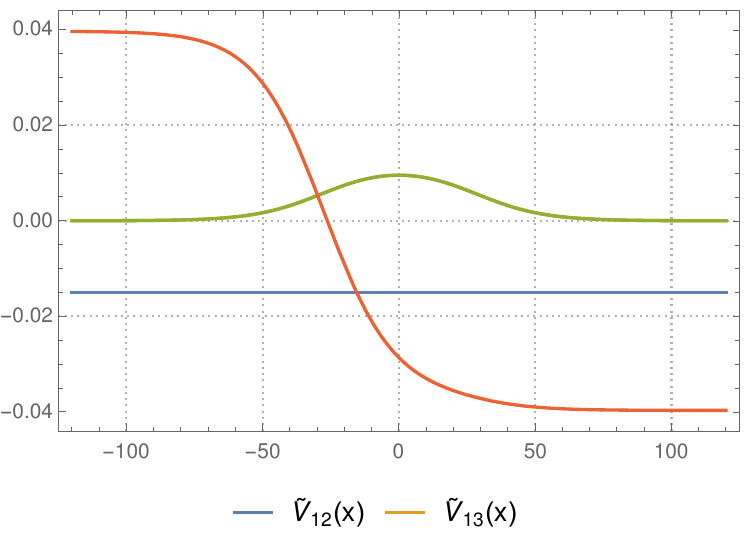}\\
			a)& b)
		\end{tabular}
		\caption{Components of the potential term of $\tilde{H}$, see (\ref{Htight}) and (\ref{vmodel2b}). In the figures, we fixed $m=0.03$, $\lambda=0.015$ on the right and $m=0.03$, $\lambda=-0.015$ on the left.   }\label{tab2}
	\end{center}
\end{table}

\section{Discussion}

The susy transformation is typically used to modify the potential term of a stationary equation. In this article, however, the transformation is approached from a different perspective. It enables the coupling of two initially non-interacting systems, such that the resulting configuration inherits properties from both systems. Specifically, the initial Hamiltonian (\ref{H}) represented two chains of atoms - the $AB$-chain and the $C$-chain that had no mutual interaction. However, the new Hamiltonian (\ref{Htight}), obtained via the susy transformation, includes interaction terms $\tilde{v}_{13}$ and $\tilde{v}_{23}$, which represent a coupling between the two atomic chains. The new Hamiltonian features a flat band with energy $\lambda$, inherited from the initially non-interacting $C$-chain. We employed this strategy to construct a solvable model of a quasi-one-dimensional chain that is locally converted into a saw-chain, as shown in Fig.\ref{fig10} and (\ref{vijcase1}).

The role of the susy transformation as a tool for coupling initially non-interacting systems aligns with the philosophy of recent work \cite{JakubskyCeleita}, where a unitary transformation was used as a "coupler" of non-interacting systems. We believe that further exploration of the susy transformation in this context as well as in the context of topological properties of the coupled systems, see e.g. \cite{Attig}, \cite{Malakar} and \cite{Datta}, is worth pursuing in future research. 

\section*{Acknowledgement}  VJ acknowledges the assistance provided by the Advanced Multiscale Materials for Key Enabling Technologies project, supported by the Ministry of Education, Youth, and Sports of the Czech Republic.  Project No. CZ.02.01.01/00/22\_008/0004558, Co-funded by the European Union.” - AMULET project.

\end{document}